\documentclass[12pt]{iopart}
\usepackage[export]{adjustbox}
\usepackage{booktabs}
\usepackage{graphicx}% Include Fig.files
\usepackage{dcolumn}% Align table columns on decimal point
\usepackage{bm}% bold math
\usepackage[caption=false]{subfig}
\usepackage{hyperref}% add hypertext capabilities

\usepackage{color}
\usepackage{soul}

\begin{document}

%\preprint{APS/123-QED}
\title[]{Effects of the Hubbard U on density functional-based predictions of BiFeO$_3$ properties}

\author{J. Kane Shenton$^{1,2,3}$, David R. Bowler$^{1,3,4}$, Wei Li Cheah$^2$}
\ead{john.shenton.10@ucl.ac.uk}

\address{$^1$
 Department of Physics \& Astronomy, University College London, Gower St, London, WC1E 6BT
}%
\address{$^2$ Institute of High Performance Computing, 1 Fusionopolis Way, \#16-16 Connexis North, Singapore 138632}
\address{$^3$ London Centre for Nanotechnology, 17-19 Gordon St, London, WC1H 0AH}%
\address{$^4$ International Centre for Materials Nanoarchitectonics (WPI-MANA), National Institute for Materials Science (NIMS), 1-1 Namiki, Tsukuba, Ibaraki 305-0044, Japan}%
% \email{david.bowler@ucl.ac.uk}

\date{\today}% It is always \today, today,

%%%%%%%%%%%%%%%%%%%%%%%%%%%%%%%%%%%%%%%%%%%%%%%%%%%%%%%%%%%%%%%%%%%%%%%%%%%%%%
\begin{abstract}

% Intro
First principles studies of multiferroic materials, such as bismuth ferrite (BFO), require methods that extend beyond standard density functional theory (DFT). The DFT+U method is one such extension that is widely used in the study of BFO.
%
% Here we...
We present a systematic study of the effects of the U parameter on the structural, ferroelectric and electronic properties of BFO.
% We find...
We find that the structural and ferroelectric properties change negligibly in the range of U typically considered for BFO (3--5 eV).
In contrast, the electronic structure varies significantly with U.
In particular, we see large changes to the character and curvature of the valence band maximum and conduction band minimum, in addition to the expected increase in band gap, as U increases.
Most significantly, we find that the $t_{2g}$/$e_{g}$ ordering at the conduction band minimum inverts for U values larger than 4 eV.
% Implications
We therefore recommend a U value of at most 4 eV to be applied to the Fe $d$ orbitals in BFO.
More generally, this study emphasises the need for systematic investigations of the effects of the U parameter not merely on band gaps but on the electronic structure as a whole, especially for strongly correlated materials.

\end{abstract}

\maketitle

%%%%%%%%%%%%%%%%%%%%%%%%%%%%%%%%%%%%%%%%%%%%%%%%%%%%%%%%%%%%%%%%%%%%%%%%%%%%%%
\section{\label{sec:intro}Introduction\protect}

The magnetoelectric multiferroic material, bismuth ferrite (BiFeO$_3$; BFO), combines a spontaneous polarisation with an antiferromagnetic ordering in a single phase, at room temperature.
This combination of properties makes BFO an interesting material for both fundamental research and a wide range of applications, from spintronics \cite{Lee2014c} to photovoltaics \cite{Ji2010,Paillard2016}.
In photovoltaic applications, the giant spontaneous polarisation ($\sim$100 $\mu C/cm^2$ \cite{Shvartsman2007,Lebeugle2007}) is thought to aid in charge separation via the bulk photovoltaic effect \cite{Fridkin2001,Ji2011}.
Ferroelectric (FE) domains are thought to further enhance the photovoltaic prospects of BFO by allowing above-band gap photovoltages across the FE domains \cite{Yang2010}, and conduction along them \cite{Seidel2009}.

Another attractive feature of the BFO system is the tunability of its properties with experimentally accessible changes to its crystal structure. A wide range of crystal structures with widely varying optoelectronic properties can be stabilised through the epitaxial strain engineering of BFO thin films \cite{Sando2014,Sando2016}.
The subtle interplay between structural and electronic degrees of freedom that underlies the tunability of the BFO system, however, make this material particularly challenging to model. For example, the weak (Dzyaloshinskii-Moriya) ferromagnetism observed in BFO cannot be captured without including the effects of spin-orbit coupling (SOC) \cite{Ederer2005}.

More generally, standard density functional theory (DFT) methods are known to have systematic failures in describing the electronic structure of materials with strongly correlated $d$ states, such as BFO. In particular, standard local density and generalised gradient approximations of the exchange-correlation (xc) functional incorrectly describe the on-site Coulomb interactions of highly localised electrons, due to erroneous electron self-interaction. This failure to describe strongly localised states exacerbates the infamous band gap problem in DFT.

A number of techniques can be used to improve the description of localised electronic states, such as self-interaction correction methods and the use of hybrid functionals.
While hybrid functionals, in particular the Heyd-Scuseria-Ernzerhof (HSE) screened hybrid functional, have been shown to accurately capture many properties of BFO, they come at a vastly increased ($\sim$50$\times$) computational cost compared to standard DFT \cite{Stroppa2010}. For simple bulk BFO, the additional cost is perfectly feasible on modern computer architectures. However, exciting developments in the study of BFO suggest that ferroelectric domains \cite{Lee2014c,Bhatnagar2013a,Seidel2009}, doping \cite{Rong2015,Wang2015w} and hetero-interfaces \cite{Rondinelli2012,Borisevich2010,Zhou2010,Yu2012,Kim2013b} with this material hold great technological promise.
The theoretical investigation of such systems requires large simulation cells, which can be prohibitively expensive using hybrid functionals.

One of the most computationally cost-effective corrections to standard DFT is the `DFT+U' method. In this method an on-site Hubbard-like correction is applied to the effective potential. Two free parameters, U and J, can be used to effectively tune the on-site Coulomb and exchange interactions respectively. In the approach of Dudarev \emph{et al.} \cite{Dudarev1998}, these are replaced by a single parameter, $\mathrm{U_{eff} = U-J}$.
While the $\mathrm{U_{eff}}$ parameter can be obtained from \emph{ab initio} calculations \cite{Pavarini:136393}, it is typically chosen semi-empirically by comparing some predicted property to the available experimental data. The property used for calibration purposes depends on the intended aim of the study, with the electronic band gap and oxidation energies \cite{Wang2006} being two of the most commonly chosen.

Careful tests are required, however, to ensure that other material properties are not adversely affected by one's choice of $\mathrm{U_{eff}}$.
For example, if $\mathrm{U_{eff}}$ is chosen such that the predicted band gap agrees with experiment, one should ensure that no significant error is introduced into the calculated lattice parameter as a result.
Previous studies of the effects of the $\mathrm{U_{eff}}$ parameter have been conducted on the structural and electronic properties of BFO. Neaton \emph{et al.} found that choosing a value of $\mathrm{U_{eff}}=4$ eV within the local spin density approximation (i.e. the LDA+U) improves the accuracy of the calculated lattice parameter, rhombohedral cell angle and the electronic band gap \cite{Neaton2005}.
For their finite-temperature study of BFO, Kornev \emph{et al.} used the scheme of Cococcioni and de Gironcoli \cite{Cococcioni2005}  to self-consistently determine the value of U within the LDA+U to be 3.8 eV \cite{Kornev2007}. However, they found that the parameters in their effective Hamiltonian were extremely sensitive to the value of U, stating that some of these parameters changed by about 20\% when U was slightly reduced from 3.8 to 3.5 eV.
Applying the U correction to the generalised gradient approximation (GGA+U), a U value of 5 eV was determined by Young \emph{et al.} to most accurately reproduce the experimental imaginary permittivity near the band gap \cite{Young2012a}.

The effects of the $\mathrm{U_{eff}}$ parameter on the crystal structure, band gap and permittivity of BFO are therefore known, but those on other electronic properties have not yet been reported for BFO as far as we are aware.
In this paper we extend the systematic study of the $\mathrm{U_{eff}}$ parameter to include the curvature and character of the band edges in BFO. We find that the electron and hole effective masses, inversely proportional to the band curvature, are highly sensitive to the chosen value of $\mathrm{U_{eff}}$ and that the ordering of the Fe $d$ orbitals at the conduction band minimum inverts for $\mathrm{U_{eff}}>4$ eV. These findings  have important implications for theoretical studies of BFO and related materials, especially in cases where the charge carrier effective masses and band character play significant roles, such as in photovoltaics \cite{Shenton}.

%%%%%%%%%%%%%%%%%%%%%%%%%%%%%%%%%%%%%%%%%%%%%%%%%%%%%%%%%%%%%%%%%%%%%%%%%%%%%%

%%%%%%%%%%%%%%%%%%%%%%%%%%%%%%%%%%%%%%%%%%%%%%%%%%%%%%%%%%%%%%%%%%%%%%%%%%%%%%
\section{\label{sec:methods}Methods\protect}

All of the results presented here are based on DFT simulations using version 5.4.1 of the Vienna \emph{ab initio} Simulation Package \cite{Kresse1993,Kresse1994,Kresse1996a,Kresse1996b} (VASP).
The calculations were carried out using the projector-augmented plane-wave method \cite{Blochl1994b,Kresse1999b}, treating explicitly 15 electrons for Bi ($5d^{10}6s^{2}6p^{3}$), 14 for Fe ($3p^{6}3d^{6}4s^{2}$), and 6 for O ($2s^{2}2p^{4}$).
\footnote{The Bi, Fe and O PAWs are dated: 6$\mathrm{^{th}}$ Sept. 2000, 2$\mathrm{^{nd}}$ Aug. 2007 and 8$\mathrm{^{th}}$ Apr. 2002 respectively}
We use a plane-wave cut-off energy of 520 eV and perform Brillouin zone integrations on a $\Gamma$-centred $\mathrm{9\times9\times9}$ Monkhorst-Pack mesh \cite{Monkhorst1976}.
The GGA xc functional parameterised by Perdew, Burke and Ernzerhof (PBE) \cite{Perdew1996} is used throughout the paper, with comparisons to the LDA and PBEsol \cite{Perdew2008} functionals where appropriate.
We apply the effective Hubbard-like correction, $\mathrm{U_{eff}}$, to the Fe $d$ orbitals using the method of Dudarev \emph{et al.} \cite{Dudarev1998}, varying the magnitude of $\mathrm{U_{eff}}$ between 0 and 8 eV.

We use the ground-state, rhombohedral BFO structure (spacegroup: $R\,3\,c$) \cite{Palewicz2007} as our model for all calculations. This phase exhibits a large spontaneous polarisation along the pseudo-cubic [111] direction ($[111]_{pc}$), primarily due to a Bi translation along this direction. This phase adopts a nearly G-type antiferromagnetic ordering \cite{Sosnowska1982b} which we approximate as exactly G-type by using a 10-atom unit cell (two formula units), with the spin on the Fe atoms alternating along the $[111]_{pc}$ direction. See the insets in Fig. \ref{fig:ueff_trans_o-fe-o} for a depiction of the structure used.

As previously mentioned, SOC has been found to be significant in describing the weak ferromagnetism in BFO \cite{Ederer2005}.
However, SOC has been found to negligibly affect the curvature and character of the band edges in $R\,3\,c$ BFO \cite{Shenton}. In particular, for the $\mathrm{U_{eff}}=4$ eV relaxed structure, the calculated absolute hole effective mass increased from 0.748 $m_0$ with SOC to 0.763 $m_0$ without SOC, where $m_0$ is the electron rest mass.
Similarly, the electron effective mass increased from 2.950 $m_0$ with SOC, to 3.017 $m_0$ without SOC.
Such differences are significantly smaller than those being investigated here, and we therefore neglect SOC hereafter.

The following procedure was repeated for each value of $\mathrm{U_{eff}}$, tested in the range $0 \leq \mathrm{U_{eff}} \leq 8$ eV:
first, a full geometry optimisation was performed in which the internal coordinates were relaxed such that all force components were less than 2 meV/\AA.
The unit cell shape and size were optimised such that all stress components were smaller than 2 MPa.
Following the geometry optimisation, an accurate self-consistent calculation was performed.

The spontaneous polarisation, $P_s$, was calculated using the Modern Theory of Polarisation (MTP) \cite{King-Smith1993,Vanderbilt1993b,Resta1994}.
We note that, according to the MTP, only \emph{differences} in polarisation are well-defined, and that bulk polarisation is best understood as a lattice of values \cite{King-Smith1993,Vanderbilt1993b,Resta1994}. In general, one needs to construct a ferroelectric switching path to resolve the ambiguity in the calculated polarisation - i.e. to find out on which branch of the polarisation lattice the calculated polarisation lies. By constructing such a switching path, we found that $P_s$ is related to our raw calculated polarisation, $P_{\mathrm{calc}}$, and the so-called quantum of polarisation, $Q$, via $P_s = P_{\mathrm{calc}} + Q/2 $. For more details on the necessity of this additional step in the context of BFO, see Ref.~\cite{Neaton2005}.

To calculate the charge carrier effective mass, $m^{*}$, we require the second derivative of the dispersion relation for a given band and location in reciprocal space. In general, effective masses are anisotropic and so a full effective mass tensor is required.
We obtain the full effective mass tensors at the valence band maximum ($\mathrm{VB_{max}}$) and conduction band minimum ($\mathrm{CB_{min}}$) following the procedure outlined in Ref.~\cite{Shenton}, using the method and code found in Ref.~\cite{Fonari2012}. Briefly, the method involves generating a fine mesh around the $k$-point of interest, calculating the energy eigenvalues at fixed, self-consistent charge density, and using a finite difference method to build up the tensor of second derivatives.
The dependence of $m^{*}$ on the mesh spacing parameter was investigated and spacings of less than 0.05 bohr$^{-1}$ were found to give consistent results \cite{Shenton}.
In order to compare the effective masses at different values of $\mathrm{U_{eff}}$ we calculated the eigenvalues of each $m^*$ tensor, which correspond to the $m^{*}$ along the principle directions, and selected the smallest eigenvalue in each case.

Several alternative approaches to estimating $m^{*}$ exist. One could, for example, focus on the curvature of a fixed band at a fixed $k$-point, for all values of $\mathrm{U_{eff}}$. This has the benefits of being more straightforward, and of isolating changes in the curvature of the chosen band from changes to the location and character of the band edges. 
Another approach would be to average $m^{*}$ across the whole of the lowest band or set of bands as was done by Hautier \emph{et al.} \cite{Hautier2013,Hautier2014}. The latter approach is of particular value in cases where the bands around the Fermi level are very flat, since in such cases multiple band extrema become energetically relevant to conduction. 
In this work we choose to focus on the curvature of $\mathrm{VB_{max}}$ and $\mathrm{CB_{min}}$, although the location may change with $\mathrm{U_{eff}}$, in order to emphasise the role of these points in determining the response of the conduction electrons/holes. With this approach, we find that abrupt changes in $m^{*}$ provides an indication of changes to the character of the band edges.

 Finally, we computed the electronic density of states.
 We note that, because the FeO$_6$ octahedra do not align with the Cartesian axes in the rhombohedral unit-cell setting, VASP fails to correctly model the fine details of the projected DOS. 
 In particular, in the rhombohedral setting, the DOS does not exhibit any of the typical splitting of the Fe $d$ orbitals that one would expect given the octahedral environment of Fe. To obtain more detailed and accurate DOS projections we converted each of the relaxed structures into their 40-atom, pseudo-cubic, unit cell setting. Although the FeO$_6$ octahedral axes still do not line up perfectly with the Cartesian axes in this setting (due to the tilting of the octahedra), a clear splitting between the projected $t_{2g}$ and $e_{g}$ states is observed in this setting, as expected by symmetry. To obtain a more accurate DOS, a finer $\mathrm{11\times11\times11}$ Monkhorst-Pack mesh was used, in addition to using the larger 40-atom unit cell.

The full VASP input files and structures are available at Ref.~\cite{Shenton2017}.
%%%%%%%%%%%%%%%%%%%%%%%%%%%%%%%%%%%%%%%%%%%%%%%%%%%%%%%%%%%%%%%%%%%%%%%%%%%%%%

%%%%%%%%%%%%%%%%%%%%%%%%%%%%%%%%%%%%%%%%%%%%%%%%%%%%%%%%%%%%%%%%%%%%%%%%%%%%%%
\section{\label{sec:results}Results and Discussion\protect}
    %-------------------------------------------%
    \subsection{Crystal structure}
    \label{subsec:crystal_struc}

    \begin{figure*}
        \centering
        \subfloat[]{\includegraphics[width=0.45\linewidth]{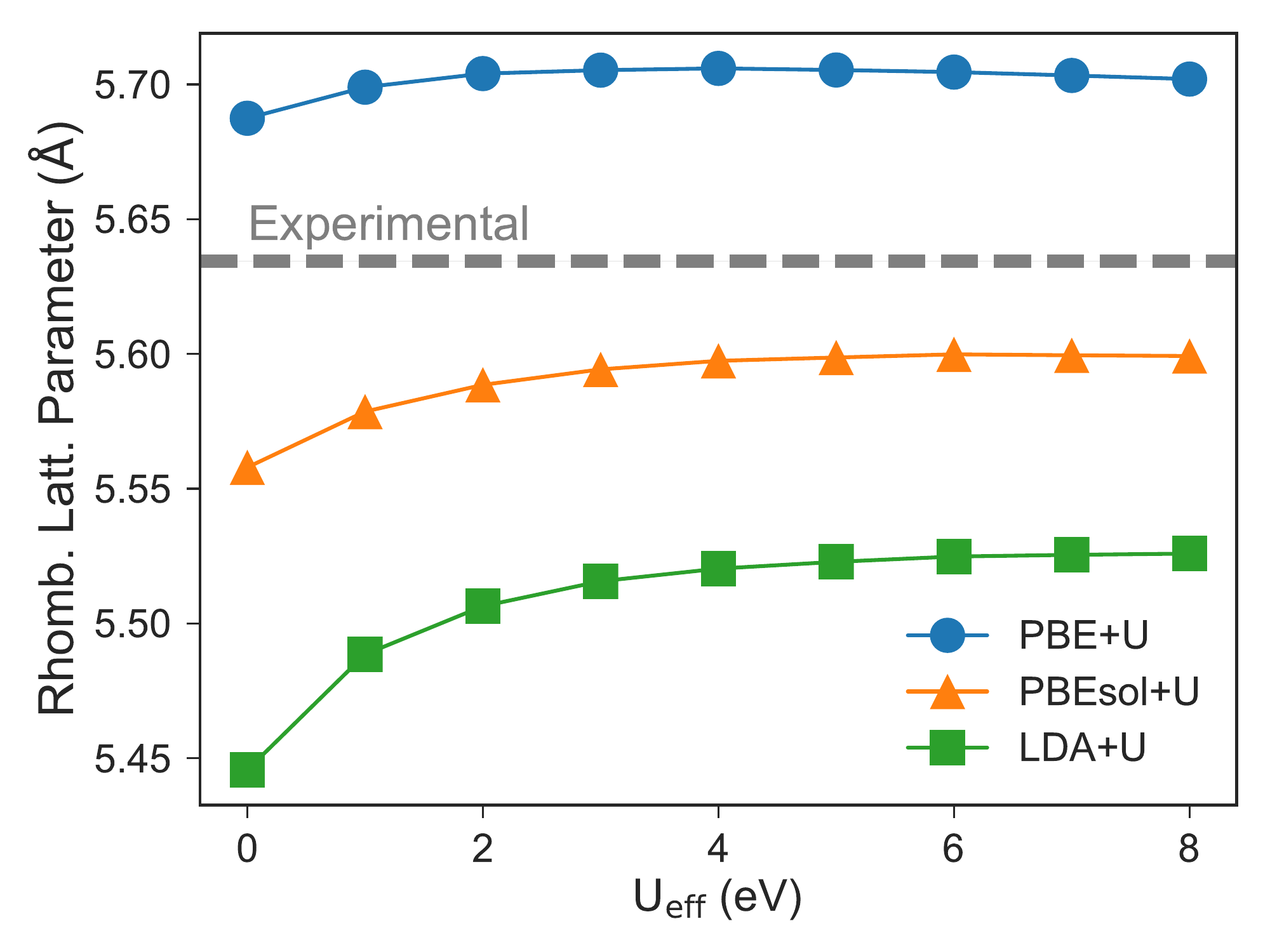}}
        \subfloat[]{\includegraphics[width=0.45\linewidth]{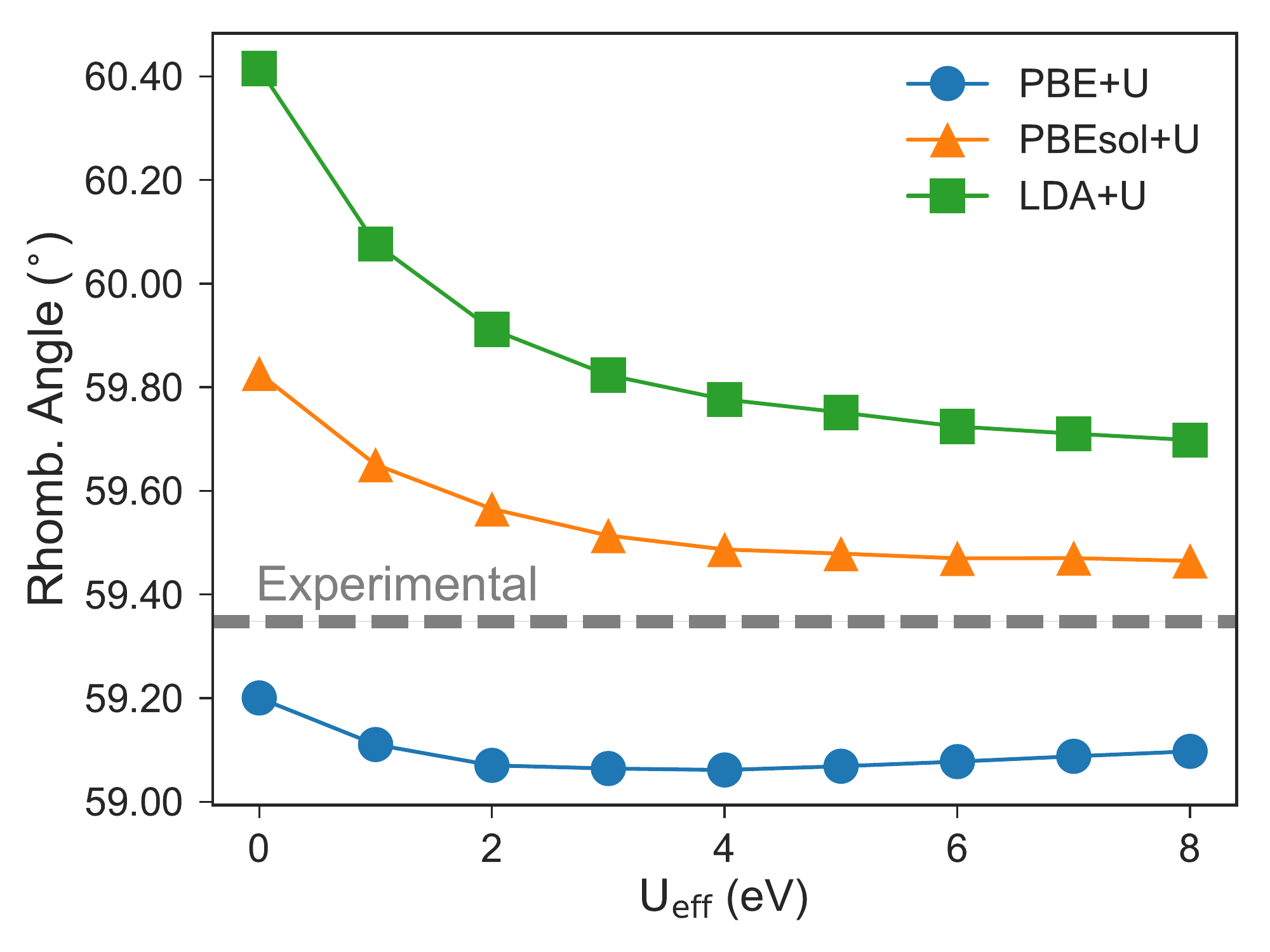}}
        \caption{Rhombohedral (a) lattice parameters and (b) cell angles as a function of $\mathrm{U_{eff}}$ across the PBE+U (blue circles), PBEsol+U (orange triangles) and LDA+U (green squares) xc functionals. The dotted `experimental' line in each comes from the (298 K) structure provided in Ref.~\cite{Palewicz2007}. The reported uncertainty of the experimental values is too small to be seen on this scale.
        }
        \label{fig:ueff_struct}
    \end{figure*}

    In Fig. \ref{fig:ueff_struct} we plot the relaxed rhombohedral lattice parameter and angle as we vary $\mathrm{U_{eff}}$ in the PBE+U xc functional. We compare the response of the PBE+U functional to that of two other commonly used xc functionals: PBEsol+U and LDA+U functionals. We find that, for all three xc functionals, increasing $\mathrm{U_{eff}}$ up to 4 eV leads to an increase in lattice parameter and a decrease in rhombohedral angle, in agreement with Neaton \emph{et al.} \cite{Neaton2005}.
    We note that an increase in lattice parameter with $\mathrm{U_{eff}}$ represents an improvement in structural accuracy for the LDA+U because of its tendency to overbind. In contrast, because the PBE functional underbinds BFO, an increase in lattice parameter constitutes a decrease in accuracy. Nevertheless, even the least accurate lattice parameter found with PBE+U ($<1.3\%$ error, occurring when $\mathrm{U_{eff}=4}$ eV) is in better agreement with the experimental lattice parameter of 5.63443(5) \AA  \cite{Palewicz2007}, than the most accurate LDA+U value ($>1.9\%$ error, occurring when $\mathrm{U_{eff}}=8$ eV). Interestingly, the PBEsol+U slightly overbinds BFO for all value of U, though performs significantly better than both the LDA+U and the PBE+U, with a lattice parameter error of $0.6\%$ occurring when $\mathrm{U_{eff}}=5\mbox{--}8$ eV.

    The error in lattice parameter can be further reduced to $\approx0.3\%$ by using the hybrid HSE functional, as reported by Stroppa \emph{et al.} \cite{Stroppa2010}. However, while very accurately reproducing the experimental lattice parameter, the authors note that this method requires around 50 times more computational time per self-consistent step than does plain PBE or PBE+U. The HSE method is therefore limited to small unit cells. For larger cells, such as those required to model defects, grain boundaries or domains, the more computationally cost-effective PBE+U method may be preferred.
    Stroppa \emph{et al.} also report a PBE relaxed structure with which our PBE results is in near perfect agreement: our calculated lattice parameter for $\mathrm{U_{eff}}=0$ eV, 5.687 \AA, agrees exactly (to all reported digits) with their results, and the rhombohedral angle differs by just $0.02^{\circ}$.

    Interestingly, for $\mathrm{U_{eff}}$ larger than 4 eV, we see the trend in lattice parameter and rhombohedral angle reverse.
    This effect may be driven by the qualitative change in the electronic structure that occurs around $\mathrm{U_{eff}}=4$ eV (as discussed in section \ref{subsec:electronic_struc}). However, further work would be needed to establish this link.

    In addition to the changes in the unit cell parameters, we observe changes in the internal coordinates of the atoms, as $\mathrm{U_{eff}}$ is increased. We are particularly interested in the two key structural transformations in $R\,3\,c$ BFO relative to the cubic perovskite structure. The first is a translation of the Bi ions \emph{along} the $[111]_{pc}$ direction; the second is an out-of-phase rotation of the FeO$_6$ octahedra \emph{about} the $[111]_{pc}$ direction ($a^-a^-a^-$ in the notation of Glazer \cite{Glazer1972a}).
    The former is the main driver for the large spontaneous polarisation in BFO, while the latter is thought to influence properties of BFO such as the charge carrier effective masses \cite{Shenton}, polar order \cite{Borisevich2010,Kim2013b} and spin state \cite{Catalan2009}.

\begin{figure*}
        \centering
        \subfloat[Bi Translation \label{sfig:ueff_trans}]{\includegraphics[width=.45\linewidth]{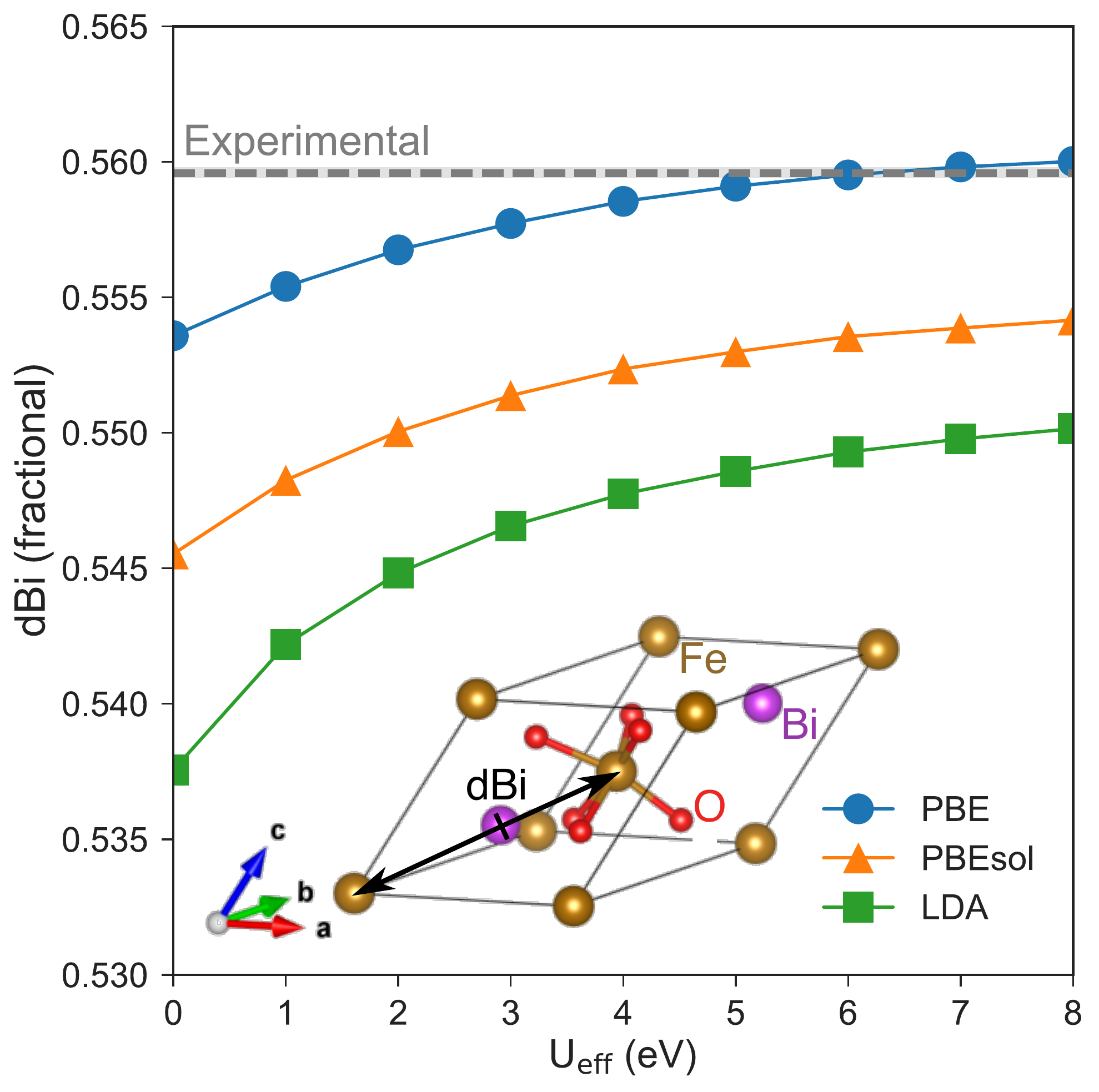}}
        %  \hfill
        \subfloat[O-Fe-O Bond Angle \label{sfig:ueff_o-fe-o}]{\includegraphics[width=0.45\linewidth]{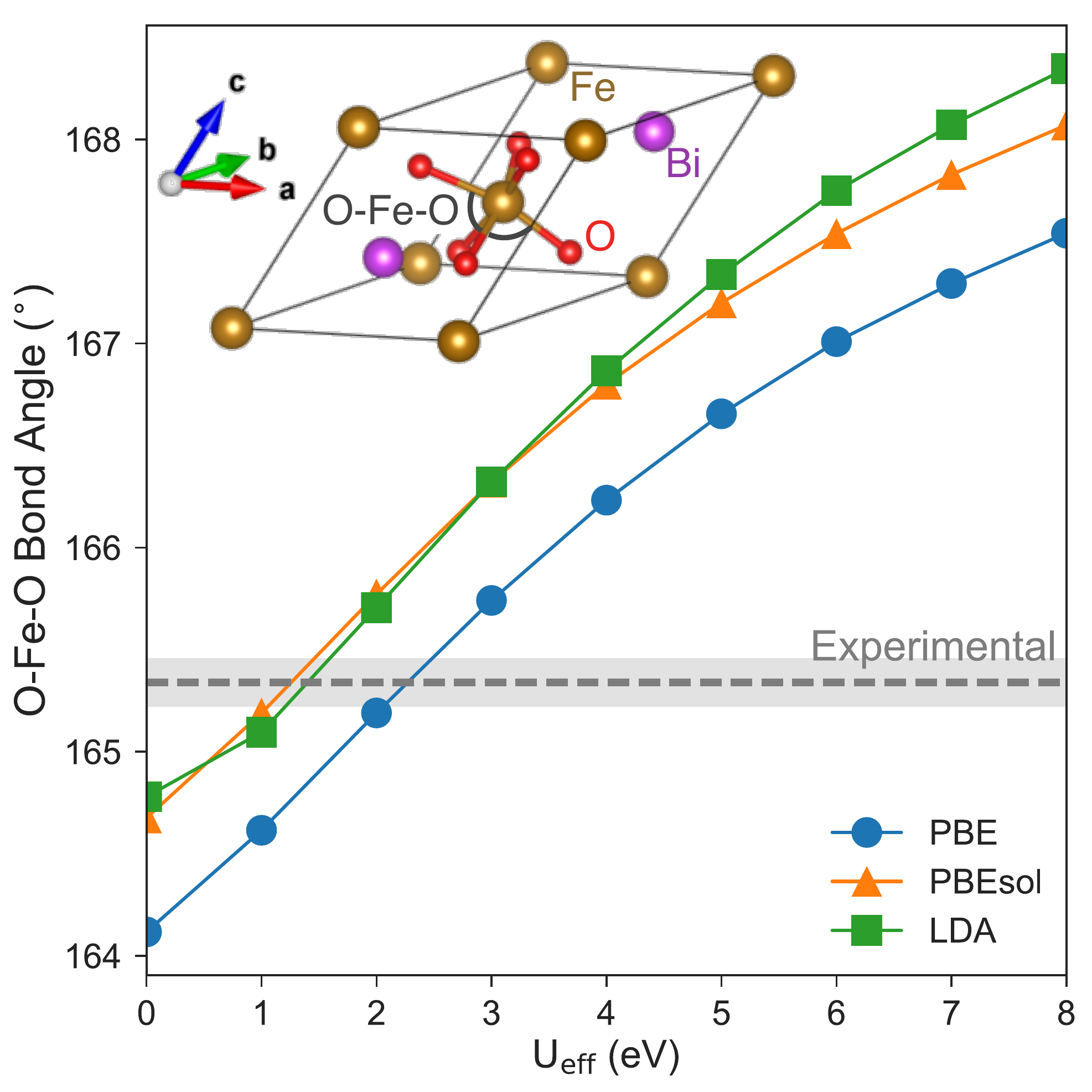}}
        \caption{Effects of $\mathrm{U_{eff}}$ on (a) position of Bi along the $[111]_{pc}$ direction as a fraction of Fe-Fe separation, and (b) O-Fe-O bond angle. Results from PBE+U (blue circles), PBEsol+U (orange triangles) and LDA+U (green squares) xc functionals are shown. The dotted `experimental' line in each is measured from the (298 K) structure provided in Ref.~\cite{Palewicz2007}, and the shaded range represents the uncertainty of those measurements. Note that in (a) this uncertainty is too small to be visible. The inset figures in each panel represent the $R\,3\,c$ structure and indicate the quantity being measured.
        }
        \label{fig:ueff_trans_o-fe-o}
    \end{figure*}

    In Fig. \ref{sfig:ueff_trans}, we represent the translation of Bi by plotting its position as a fraction of Fe-Fe separation along the $[111]_{pc}$ direction. In the perfect cubic perovskite structure, a Bi atom would lie exactly halfway between two Fe atoms in the $[111]_{pc}$ direction (i.e. dBi$=0.5$ as defined in Fig. \ref{sfig:ueff_trans}). Compared with the experimental fractional translation, $\mathrm{dBi_{exp}}=0.55958(18)$, we see an improvement in the description of the translation of Bi as $\mathrm{U_{eff}}$ is increased.
    At $\mathrm{U_{eff}}=6$ eV, dBi most closely matches that found in experiment.
    Note that we compare the fractional displacement of Bi (as opposed to absolute displacements), in order to take into account the changing lattice parameters at each value of $\mathrm{U_{eff}}$.

    In the $R\,3\,c$ structure of BFO, the FeO$_6$ octahedra are rotated about $[111]_{pc}$ by $\sim$14$^{\circ}$.
    We find that the angle of rotation changes by $\approx 0.4^{\circ}$ when varying $\mathrm{U_{eff}}$ between 0 and 8 eV. Given that the angle of this rotation found in experiment spans the range 11--14$^{\circ}$ \cite{Moreau1971,Kubel1990,Megaw1975}, we conclude that the change in octahedral rotation due to $\mathrm{U_{eff}}$ is negligible.

    Another manifestation of the distortion of $R\,3\,c$ BFO with respect to the cubic perovskite structure is the deviation of the O-Fe-O octahedral angle from 180$^{\circ}$.
    We find that the dependence of this angle on $\mathrm{U_{eff}}$ is also small, increasing from $164.1^{\circ}$ to $167.5^{\circ}$ as $\mathrm{U_{eff}}$ increases from 0 to 8 eV. In Fig. \ref{sfig:ueff_o-fe-o} we show this increase in O-Fe-O bond angle towards 180$^{\circ}$ as a function of $\mathrm{U_{eff}}$.
    Nevertheless, all of the O-Fe-O angles predicted here are in reasonably good agreement with the experimental angle of $165.34(12)^{\circ}$, determined by high-resolution neutron diffraction at 298 K \cite{Palewicz2007}.

%-------------------------------------------%
\subsection{Polarisation}
\label{subsec:polarisation}
%-------------------------------------------%

From the observed changes to the lattice geometry with varying $\mathrm{U_{eff}}$, one might expect the spontaneous polarisation $P_s$ to be affected.
The increased Bi translation with $\mathrm{U_{eff}}$ would suggest an \emph{increase} in $P_s$ with increasing $\mathrm{U_{eff}}$ since the dipole moment per unit cell increases. However, as we have found the lattice parameter (and hence unit cell volume) increases with $\mathrm{U_{eff}}$, we may expect an overall \emph{decrease} in $P_s$ as $\mathrm{U_{eff}}$ increases (recall that polarisation is inversely proportional to unit cell volume).
    We note that the structural changes due to $\mathrm{U_{eff}}$ may affect both the ionic and the electronic contributions to $P_s$. At the same time, independent of any structural changes, increasing $\mathrm{U_{eff}}$ itself may result in additional changes to the electronic contribution.
    We distinguish between structural and purely electronic effects by calculating $P_s$ both for the DFT relaxed structures (`relaxed'), and for a chosen fixed structure (`fixed') in which we only vary $\mathrm{U_{eff}}$. The fixed structure used for this purpose was that relaxed at $\mathrm{U_{eff}=4}$ eV.

\begin{figure}
        \centering
        \includegraphics[width=0.5\linewidth]{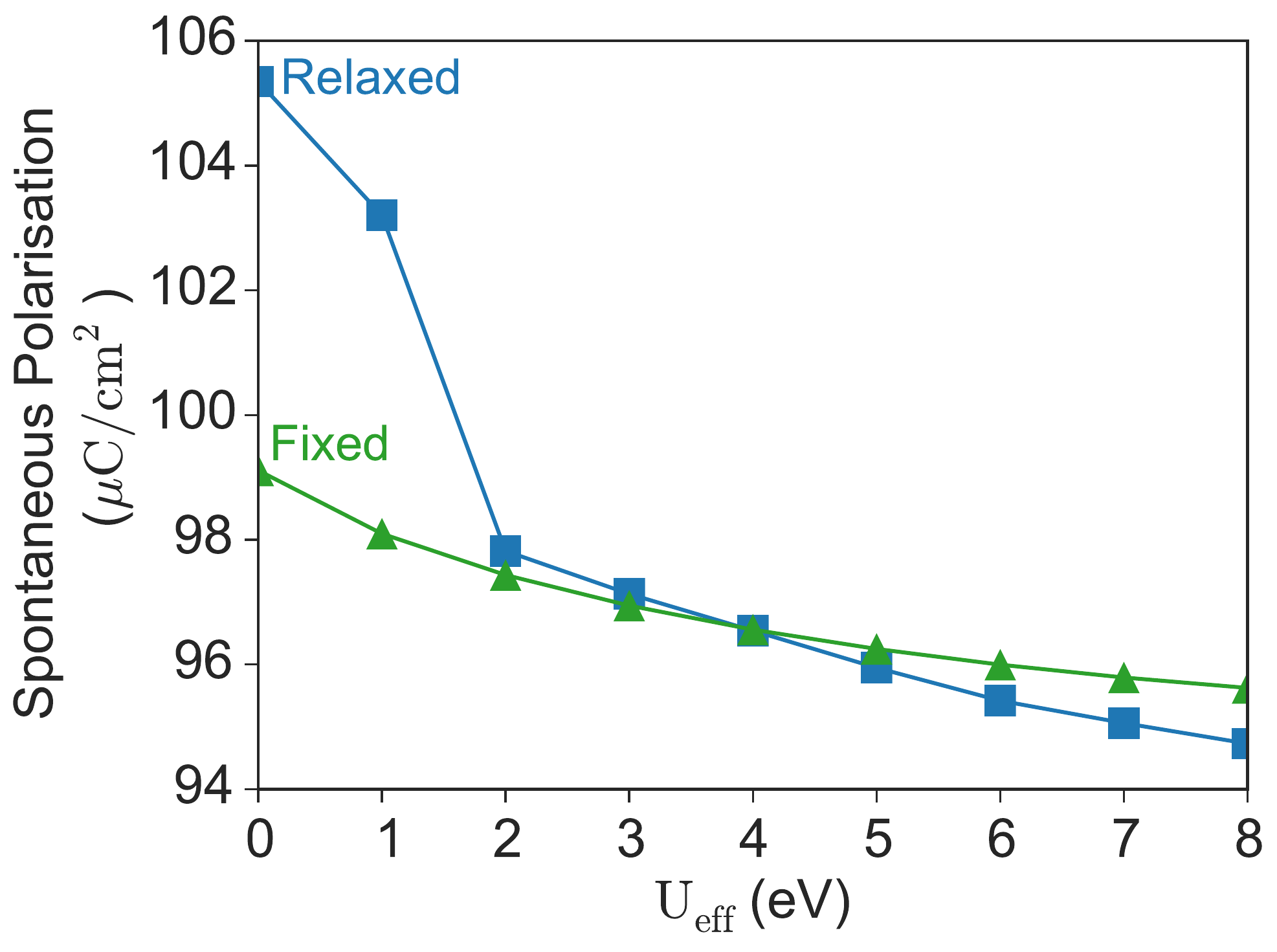}
        \caption{Variation in calculated spontaneous polarisation as a function of $\mathrm{U_{eff}}$. We compare two cases: one in which we relax the structure fully for each value of $\mathrm{U_{eff}}$ (blue squares), and the second, in which we keep the structure fixed (green triangles) to that of $\mathrm{U_{eff}=4}$ eV.
        }
        \label{fig:ueff_pol}
    \end{figure}

    In Fig. \ref{fig:ueff_pol} we plot the variation in $P_s$ with increasing $\mathrm{U_{eff}}$ for both the relaxed and fixed set of structures.
    There are two regimes present in Fig. \ref{fig:ueff_pol}: a sharp decrease in $P_s$ with respect to $\mathrm{U_{eff}}$, followed by much weaker dependence.
    The first regime, $\mathrm{U_{eff}}\leq 2$ eV, can be explained by the sharp increase in lattice parameters.
    We note that the increase in lattice parameters dominates over the increase in Bi translation along $[111]_{pc}$ that would otherwise suggest an \emph{increase} in $P_s$.
    The second regime, $\mathrm{U_{eff}}>2$ eV, in which we see a weaker dependence of $P_s$ on $\mathrm{U_{eff}}$, is dominated by changes only in the electronic structure, as the relaxed and fixed structure cases have the same dependence on $\mathrm{U_{eff}}$ beyond 2 eV.

    As with the changes in lattice parameter and angles with $\mathrm{U_{eff}}$, the most significant change in $P_s$ occurs between a $\mathrm{U_{eff}}$ of 0 and 2 eV, with only minor changes thereafter.
    Since the typical values of $\mathrm{U_{eff}}$ chosen for Fe $d$ orbitals in BFO lie between 3 and 5 eV \cite{Neaton2005, Kornev2007, Chen2011a, Chu2012, Young2012a, Tutuncu2008, Rong2015, Jin2015, Wang2015w}, the accuracy of calculated unit cell parameters and $P_s$ depends more on the choice to use the PBE+U method at all, rather than on the particular value of $\mathrm{U_{eff}}$ one chooses.
    Thus, within the range of $\mathrm{U_{eff}}$ usually considered in the context of BFO, we conclude that the crystal structure varies negligibly with $\mathrm{U_{eff}}$.

    %-------------------------------------------%
    \subsection{Electronic structure}
    \label{subsec:electronic_struc}

\begin{figure}
        \centering
        \includegraphics[width=0.5\linewidth]{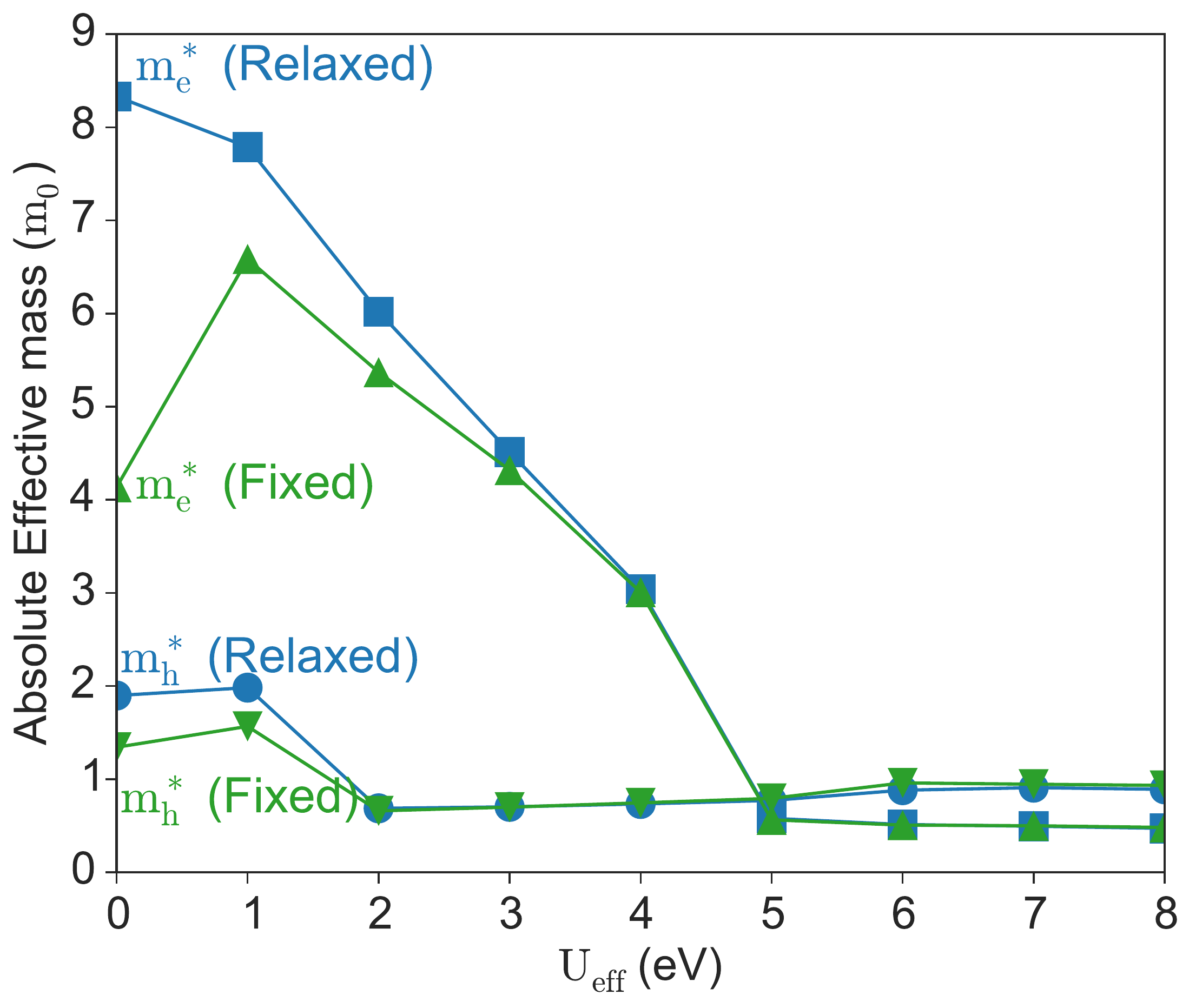}
        \caption{Absolute charge carrier effective mass versus $\mathrm{U_{eff}}$. Blue squares and circles respectively represent the electron and hole effective masses for the DFT relaxed structures. Green triangles pointing up and down represent, respectively, the electron and hole effective masses for the fixed structure (fixed to the $\mathrm{U_{eff}}=4$ eV structure). $m^{*}_{e}$ and $m^{*}_{h}$ are the electron and hole effective masses, in units of the electron rest mass, $m^{*}_{0}$.
        }
        \label{fig:ueff_mass}
    \end{figure}

\begin{figure*}
        \centering
        \begin{tabular}{cccc}
            % \centering
            \multicolumn{2}{l}{
            \subfloat{\includegraphics[width=0.15\linewidth]{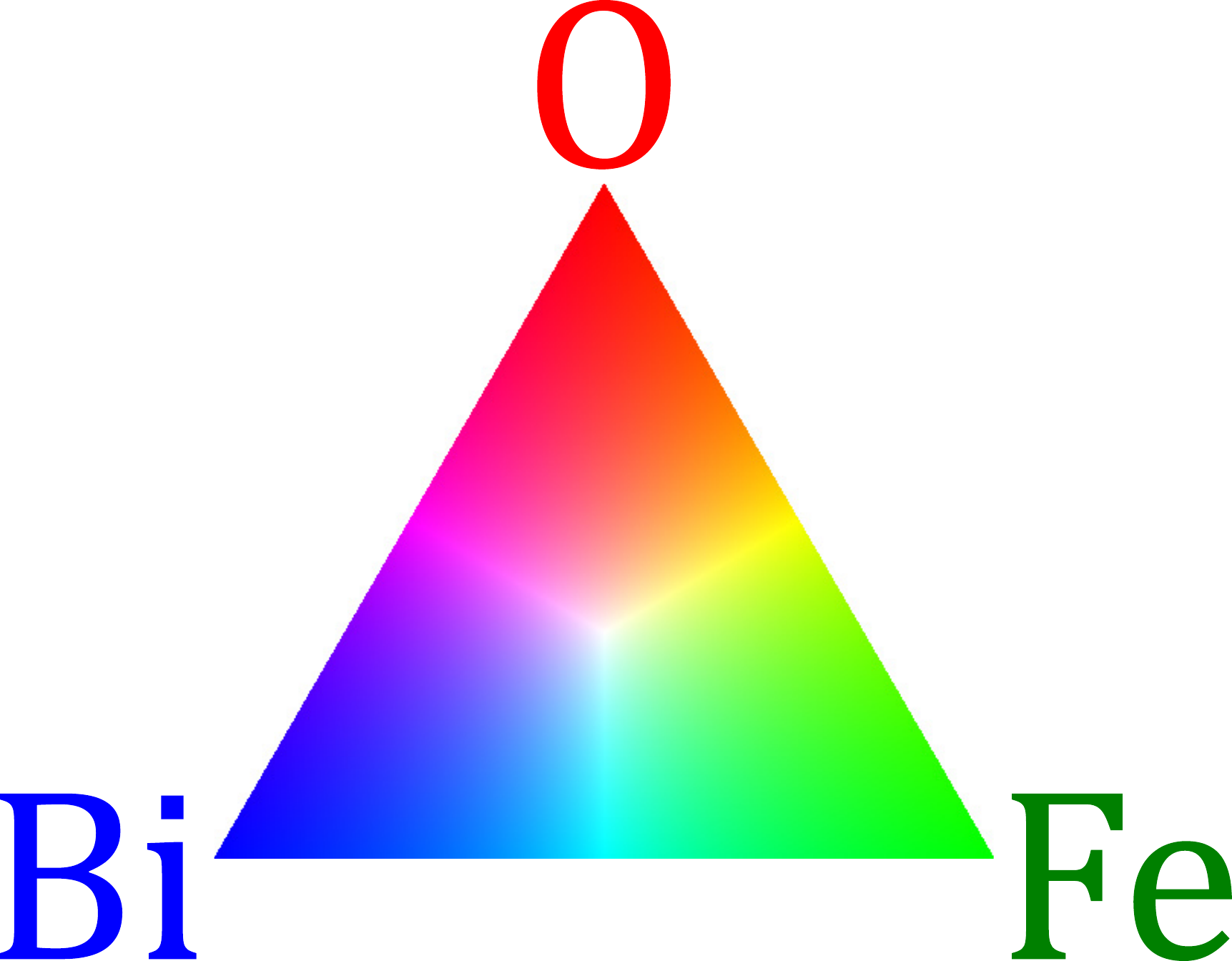}}} \\
            % % Big figures:
            \multicolumn{4}{c}{
            \subfloat{\includegraphics[width=0.9\linewidth]{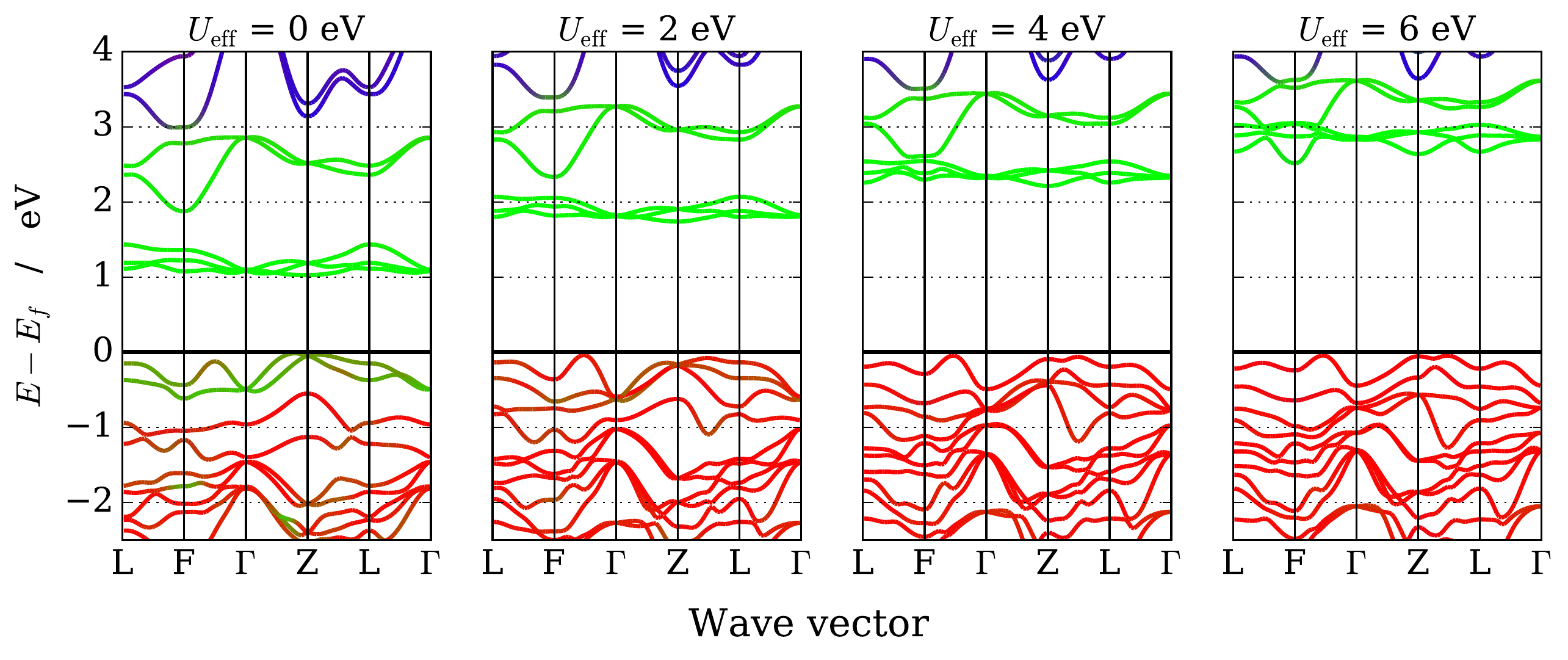}}} \\

            \multicolumn{2}{l}{
            \subfloat{\includegraphics[width=0.35\linewidth]{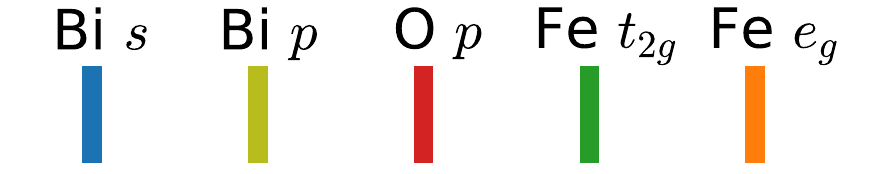}}} \\

            \multicolumn{4}{c}{
            \subfloat{\includegraphics[width=0.9\linewidth]{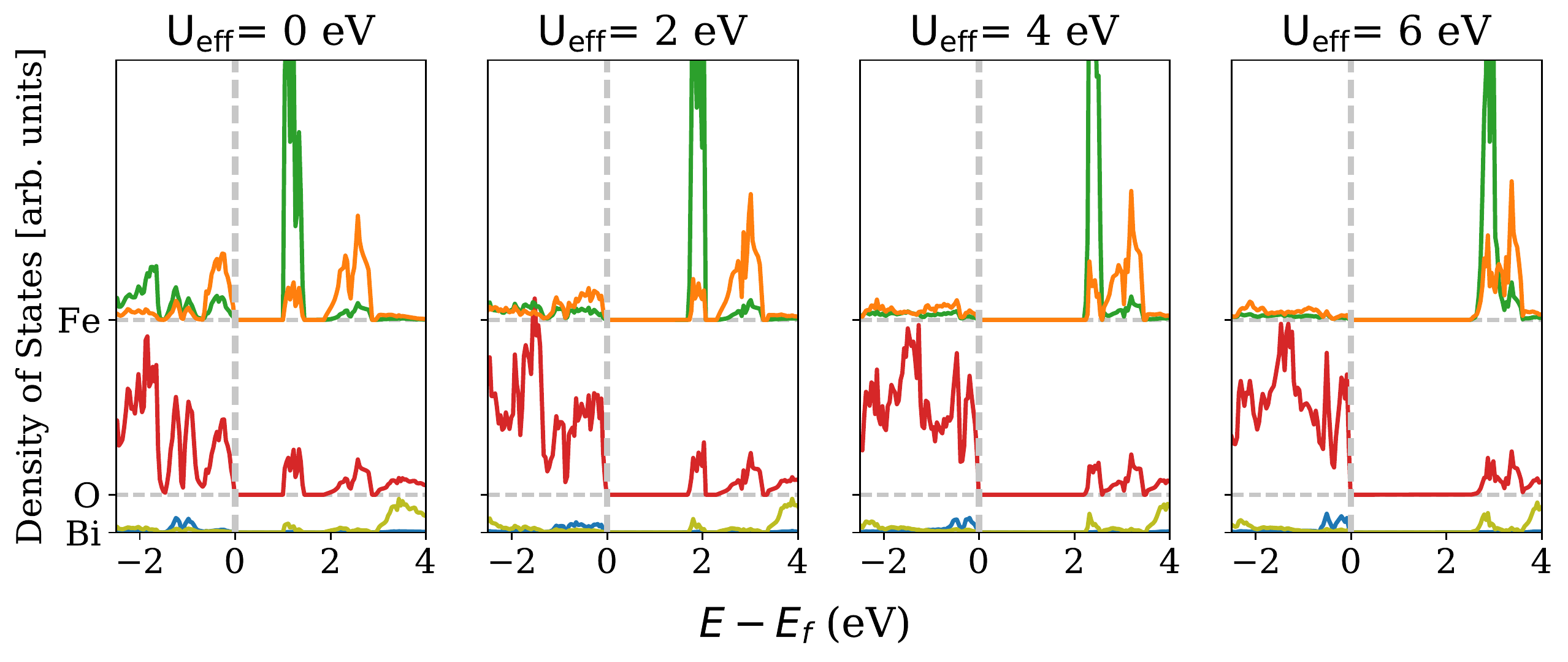}}} \\

        \end{tabular}
        \caption{Variation in projected bands (top row) and density of states (bottom row) as a function of $\mathrm{U_{eff}}$.
        The bands are coloured, at each $k$-point, based on wavefunction projections onto the elements. The contributions from O, Fe and Bi are represented on a normalised colourspace by red, green and blue respectively as shown by the colour triangle.
        The DOS is calculated in the pseudo-cubic setting in order to obtain accurate projections onto the selected atomic orbitals (indicated by the coloured vertical lines).
        Note that, since BFO adopts a G-type anti-ferromagnetic ordering, the spin up and spin down contributions are symmetrical. Here we are plotting the sum over both spin channels and over all atoms of each species.
           }
        \label{fig:ueff_pbands}
    \end{figure*}

    The influence of $\mathrm{U_{eff}}$ on the band character and curvature, often neglected in studies using DFT+U, will be the focus of this section. We quantify the curvature at the band extrema by calculating the charge carrier effective masses (Fig. \ref{fig:ueff_mass}), and represent the character using projected band structures and DOS (Fig. \ref{fig:ueff_pbands}).

    % --------------------%
    % --------------------%

    As with the polarisation, changes to the effective masses with $\mathrm{U_{eff}}$ can be broken down into structural contributions (by relaxing the structure at each $\mathrm{U_{eff}}$) and purely electronic ones (by keeping the structure fixed in each case).
    In Fig. \ref{fig:ueff_mass} we plot the electron and hole effective masses as a function of $\mathrm{U_{eff}}$ for both the relaxed and fixed structures.

    For the relaxed structures we see a large reduction in the electron effective mass $m^{*}_{e}$: from 8.3 $m_0$ for $\mathrm{U_{eff}}=0$ eV to 0.6 $m_0$ for $\mathrm{U_{eff}}=5$ eV, indicating an increase in curvature at the $\mathrm{CB_{min}}$ with increasing $\mathrm{U_{eff}}$. Between a $\mathrm{U_{eff}}$ of 5 and 8 eV we see little ($\sim$0.1 $m_0$) further change in $m^{*}_{e}$.
    The curvature of the $\mathrm{VB_{max}}$ also increases with $\mathrm{U_{eff}}$, though most of the change occurs between a $\mathrm{U_{eff}}$ of 0 and 2 eV. The hole effective mass $m^{*}_{h}$ decreases from 1.9 $m_0$ for $\mathrm{U_{eff}}=0$ eV to 0.7 $m_0$ for $\mathrm{U_{eff}}=2$ eV.

    For the cases in which the structure was kept fixed to the $\mathrm{U_{eff}}=4$ eV relaxed structure, we see a similar dependence of $m^{*}$ on $\mathrm{U_{eff}}$.
    This similarity in trend between the relaxed and fixed structure cases indicates that changes in $m^{*}_{h}$ and $m^{*}_{e}$ with $\mathrm{U_{eff}}$ are dominated by changes purely to the electronic structure.
    The notable exception to this similarity is the $m^{*}_{e}$ calculated for the fixed structure when $\mathrm{U_{eff}}$ = 0 eV ($m^{*}_{e}=4.1$ $m_0$).
    The reason for this anomaly is a change in the location of the $\mathrm{CB_{min}}$ relative to all other $\mathrm{U_{eff}}<5$ eV cases. The $\mathrm{CB_{min}}$ for the anomalous result lies between $\Gamma$ and $Z=[\frac{1}{2},\frac{1}{2},\frac{1}{2}]$, rather than exactly at $Z$ as it is for the other $\mathrm{U_{eff}}<5$ eV cases.
    We attribute the change in location to the effective tensile strain resulting from using the $\mathrm{U_{eff}}=4$ eV relaxed geometry; see Ref. \cite{Shenton} for more details on the effects of strain on $m^{*}$ in BFO.

%  Figs using tabular:
    % --------------------%

    In order to explain the observed changes in $m^{*}$ with $\mathrm{U_{eff}}$, we investigate changes to the band character as $\mathrm{U_{eff}}$ increases. We represent the band character using the projected band structure shown in Fig. \ref{fig:ueff_pbands}. The contributions from O, Fe and Bi to each band at each $k$-point are represented on a normalised colourspace by red, green and blue respectively. Similar figures comparing the projected bands of the PBE+U, PBEsol+U and LDA+U xc functionals can be found in Fig. S6 of the SI \cite{Shenton2017}. We find only minor differences between the band structures of these three xc functionals.
    We also present the projected density of states (DOS) in this figure, in order to resolve the contributions from individual orbitals.

    Beginning with the character of the valence bands, we make the following observations.
    A change in the character of the topmost valence bands, occurring between a $\mathrm{U_{eff}}$ of 0 and 2 eV, is clear from the colour change in the bands. The change from a mix of red and green at $\mathrm{U_{eff}}=0$ eV to almost pure red at $\mathrm{U_{eff}}=2$ eV in the topmost valence band (VB) indicates a reduction in the Fe-O hybridisation, leaving O to dominate the $\mathrm{VB_{max}}$. From the projected DOS we resolve these contributions further: at $\mathrm{U_{eff}}=0$ eV, the top of the VB is made up of a hybridisation of O $p$ and Fe $e_g$ states; the character of these bands changes to primarily O $p$, with minor contributions from Bi $s$ and Fe $e_{g}$ states above a $\mathrm{U_{eff}}$ of about 2 eV.
    The change in the character corresponds to the decrease in $m^*_h$ at $\mathrm{U_{eff}}=2$ eV.
    Given that the lattice vectors change most significantly in the 0 to 2 eV range of $\mathrm{U_{eff}}$, we might expect that the changes to the crystal structure of BFO are driving this shift in band character. However, as we saw in Fig. \ref{fig:ueff_mass}, the associated pattern in $m^{*}_{h}$ is similar in both the relaxed and fixed structure cases, indicating that the change is dominated by purely electronic effects. That this is not purely a structural effect is confirmed by observing the same shift in character in the projected bands and DOS for the fixed structure calculations, which can be found in Fig. S4 \cite{Shenton2017}.

    In addition to the decrease in Fe contributions to the $\mathrm{VB_{max}}$, the projected DOS shows that a Bi $s$ antibonding peak moves up in energy from around 1.5 eV below the $\mathrm{VB_{max}}$ for $\mathrm{U_{eff}}$ of 0 eV, to the $\mathrm{VB_{max}}$ itself for $\mathrm{U_{eff}}\geq 2$ eV.
    The presence of a small Bi $s$ contribution to the $\mathrm{VB_{max}}$ is consistent with the HSE hybrid functional results by Stroppa and Picozzi \cite{Stroppa2010}.
An experimental study comparing $\mathrm{VB_{max}}$ energies of BFO, Bi$_2$O$_3$ and Fe$_2$O$_3$ also proposes a non-negligible contribution from the Bi $s$ states, as well as Fe $d$ states, to the {$\mathrm{VB_{max}}$} of BFO \cite{Li2013}.
The findings of these two previous works are better reflected in our calculated electronic structures of the $\mathrm{VB_{max}}$ for $\mathrm{U_{eff}}\geq 2$ eV.

    % --------------------%
    \begin{figure*}
        \centering
        \begin{tabular}{ccc}
             \subfloat[LUKS$_Z$ $\mathrm{U_{eff}=0}$ eV ]{\includegraphics[width=0.28\linewidth]{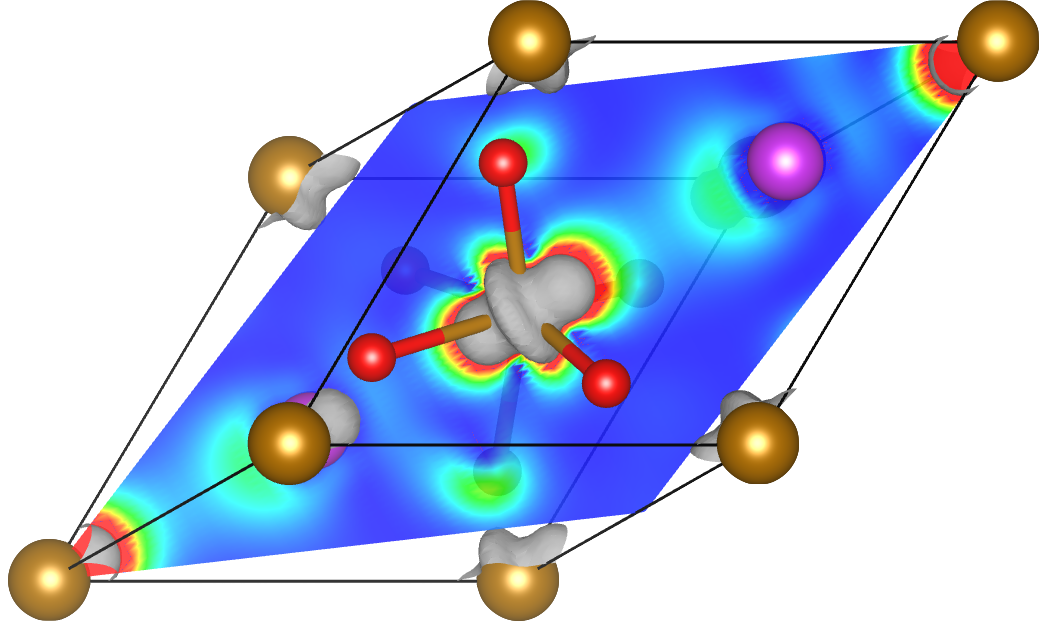}} &
             \subfloat[LUKS$_Z$ $\mathrm{U_{eff}=4}$ eV ]{\includegraphics[width=0.28\linewidth]{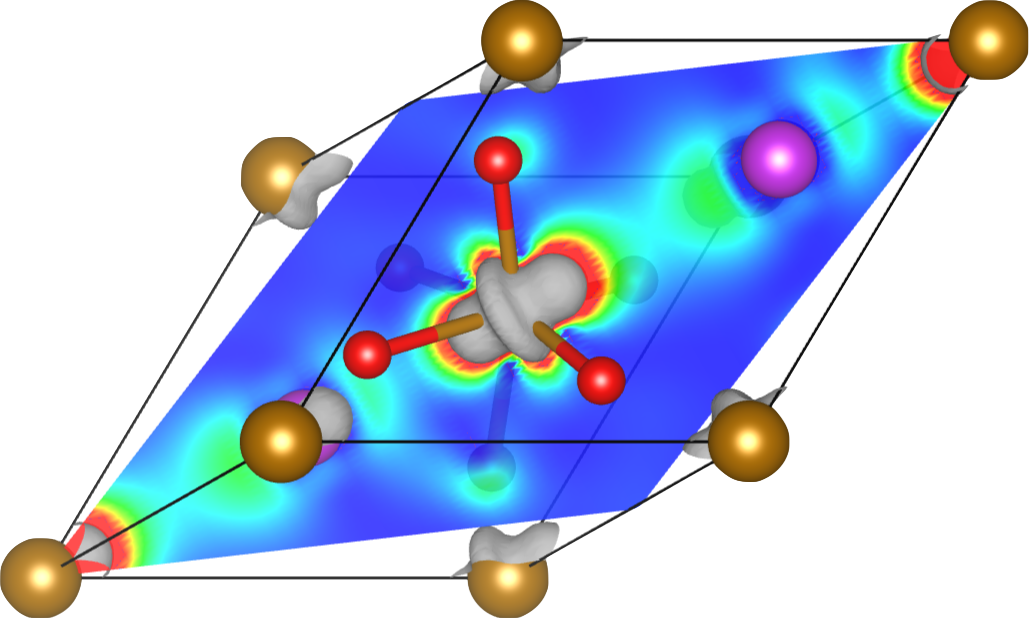}} &
             \subfloat[LUKS$_Z$ $\mathrm{U_{eff}=6}$ eV ]{\includegraphics[width=0.28\linewidth]{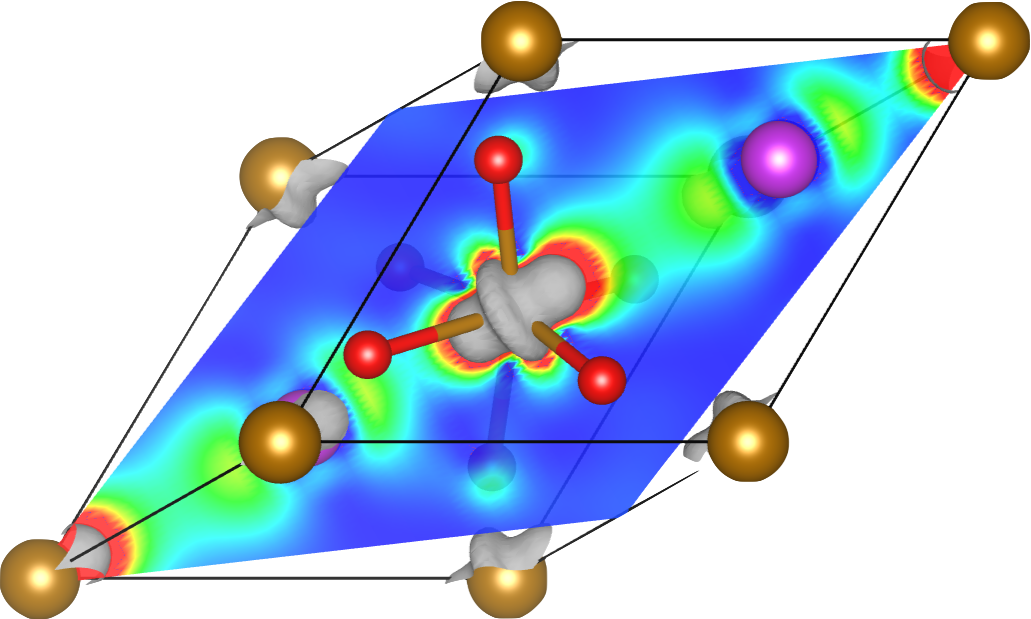}} \\
             \subfloat[LUKS$_F$ $\mathrm{U_{eff}=3}$ eV ]{\includegraphics[width=0.28\linewidth]{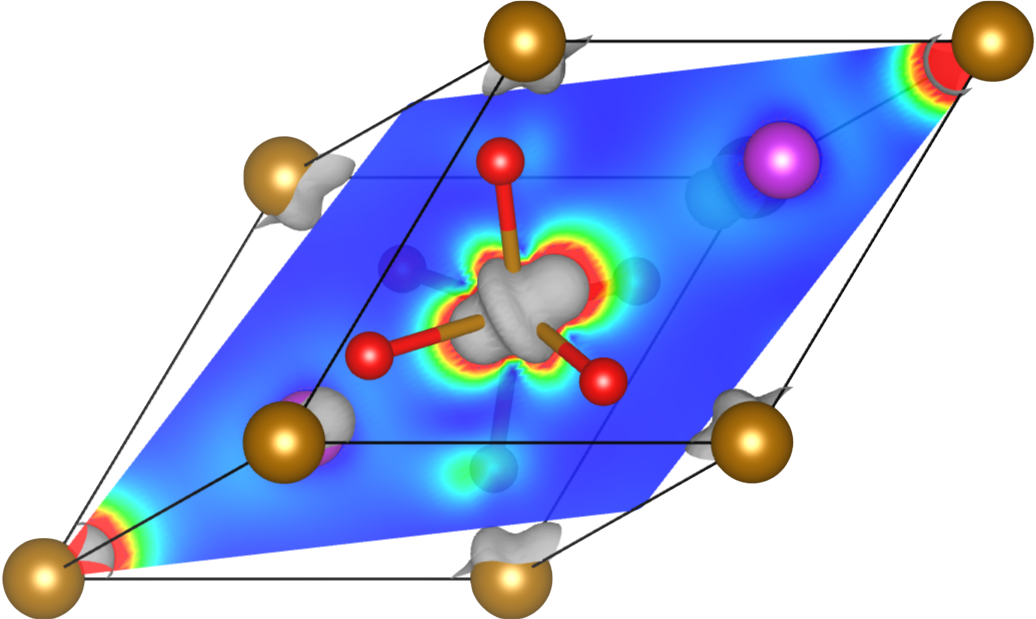}} &
             \subfloat[LUKS$_F$ $\mathrm{U_{eff}=4}$ eV ]{\includegraphics[width=0.28\linewidth]{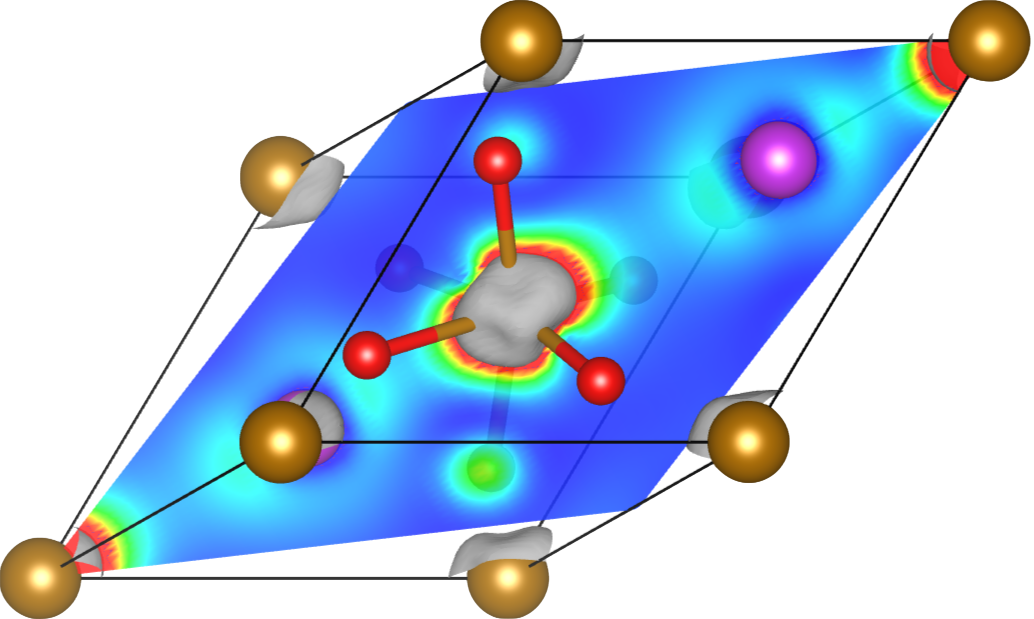}} &
             \subfloat[LUKS$_F$ $\mathrm{U_{eff}=5}$ eV ]{\includegraphics[width=0.28\linewidth]{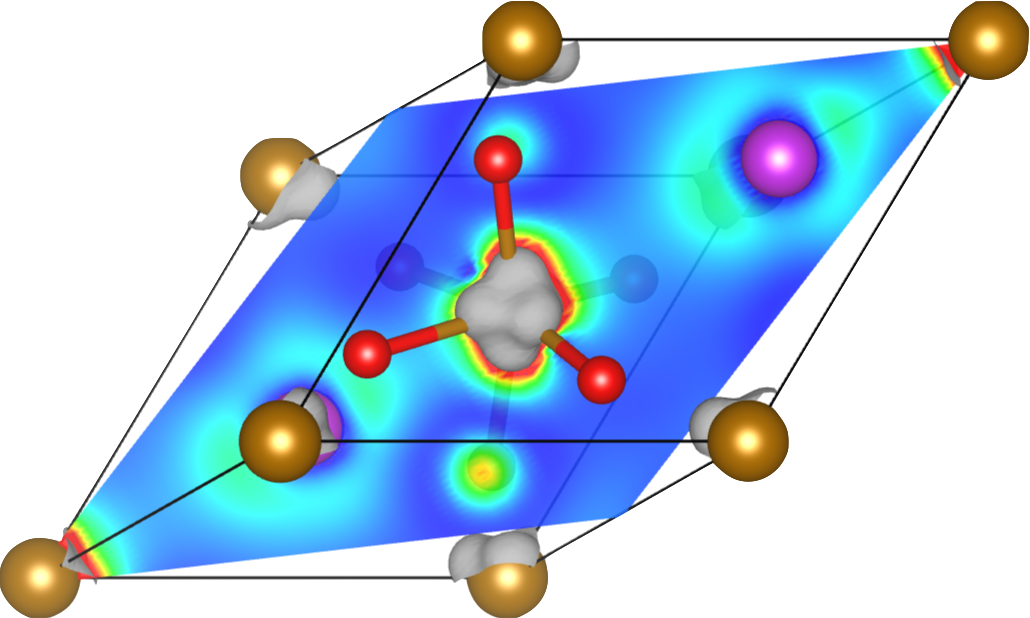}} \\
             \includegraphics[width=0.1\linewidth]{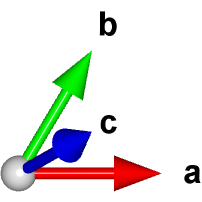} &
             \includegraphics[width=0.4\linewidth]{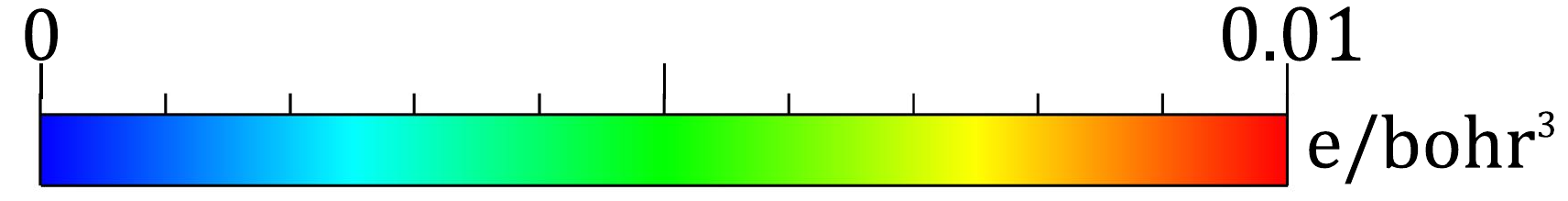} &
             \includegraphics[width=0.15\linewidth]{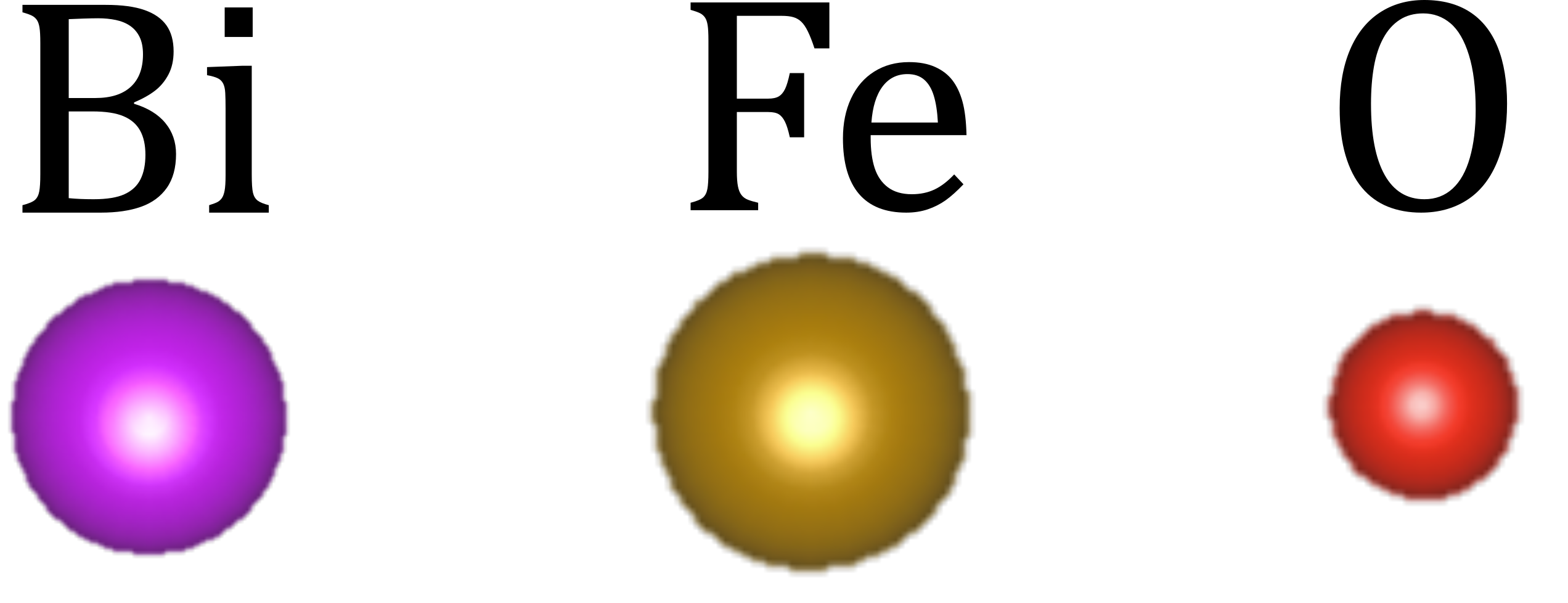}
        \end{tabular}
        \caption{Lowest unoccupied KS orbitals (LUKS) at the $Z$ (a-c) and $F$ (d-f) $k$-points for various values of $\mathrm{U_{eff}}$. The KS orbitals are represented by a gray 0.02 e/bohr$^{3}$ isosurface. We plot the charge density from the KS orbitals in a $(1\,1\,\bar{2})$ plane, with colour saturation levels indicated by the colour bar.
           }
        \label{fig:ueff_ks}
    \end{figure*}

    From the projected conduction bands (Fig. \ref{fig:ueff_pbands}), we see little change in the \emph{elemental} contributions to the $\mathrm{CB_{min}}$; Fe dominates the $\mathrm{CB_{min}}$ for all $\mathrm{U_{eff}}$ investigated here.
    The band structures do indicate, however, a change in the relative \emph{orbital} contributions to the $\mathrm{CB_{min}}$.
    We might expect the three lowest unoccupied bands to be Fe $t_{2g}$ in character and the two next unoccupied bands to be Fe $e_{g}$, based on the octahedrally coordinated Fe. Indeed, we see at the $\Gamma$-point that the five Fe $d$ bands form neatly into distinct triply and doubly degenerate sets. Detailed analysis of the crystal-field splitting in this system could be achieved using Wannier functions as in Ref.~\cite{Scaramucci2015}, though this lies beyond the scope of the present work.
    The designation of the five green (Fe) bands into $t_{2g}$ and $e_{g}$ groups in order of increasing energy is supported by the projected DOS, in which we see the energy difference between the $e_g$ and $t_{2g}$ manifolds decrease with an increase in $\mathrm{U_{eff}}$.
    For values of $\mathrm{U_{eff}}$ greater than 4 eV however, Fig. \ref{fig:ueff_pbands} suggests that one of the two Fe $e_{g}$ bands dips below the three Fe $t_{2g}$ bands. That is, above a $\mathrm{U_{eff}}$ of 4 eV, the character of the $\mathrm{CB_{min}}$ transitions from Fe $t_{2g}$ to Fe $e_{g}$.

    To investigate the shift in orbital character from $t_{2g}$ to $e_{g}$ further, we plot the lowest unoccupied Kohn-Sham (KS) orbitals in  Fig. \ref{fig:ueff_ks}.
    Because the $\mathrm{CB_{min}}$ is located at $Z=[0.5, 0.5, 0.5]$ for $\mathrm{U_{eff}}\leq4$ eV and at $F=[0.5, 0.5, 0.0]$ for $\mathrm{U_{eff}}>4$ eV, we plot the KS orbitals at each of these locations, for relevant values of $\mathrm{U_{eff}}$. To visualise some of the more subtle changes, we plot the charge density from the KS orbitals in a ($1\,1\,\bar{2}$) plane, in addition to the isosurface. These plots highlight two distinct effects that $\mathrm{U_{eff}}$ has on the character of the $\mathrm{CB_{min}}$.

    Firstly, at the $Z$ point (which is the $\mathrm{CB_{min}}$ for $\mathrm{U_{eff}}<5$ eV), there is a gradual increase in hybridisation between the Fe $t_{2g}$ and Bi $p$ orbitals along the $[111]_{pc}$ direction as $\mathrm{U_{eff}}$ increases, evident from the increase in intensity between Fe and Bi on the ($1\,1\,\bar{2}$) plane.
    This increase in overlap between Fe and Bi states contributes to the decrease in $m^{*}_{e}$ between a $\mathrm{U_{eff}}$ of 0 and 4 eV.

    Secondly, at the $F$ point, we see a transition from $t_{2g}$ to $e_{g}$ character from the KS isosurface plot as $\mathrm{U_{eff}}$ increases from 3 eV to 5 eV, as expected from the band structures. Additionally we see a slight increase in intensity around the Bi and O atoms on the ($1\,1\,\bar{2}$) plane.
    The transition to $e_{g}$ character is associated with a further decrease in $m^{*}_{e}$, possibly due to the increased overlap with O $p$ states.
    Above 5 eV there is little change in the KS orbitals (see Fig. S3 of the SI \cite{Shenton2017}), and correspondingly, we see little change in $m^{*}_{e}$.

    The significant shift in the location and character of the $\mathrm{CB_{min}}$ for $\mathrm{U_{eff}}>4$ eV suggests that $\mathrm{U_{eff}}=4$ eV be taken as a maximum, at least in cases for which the character and curvature of the $\mathrm{CB_{min}}$ plays a significant role.

    % --------------------%
    \begin{figure}
        \centering
        \includegraphics[width=0.5\linewidth]{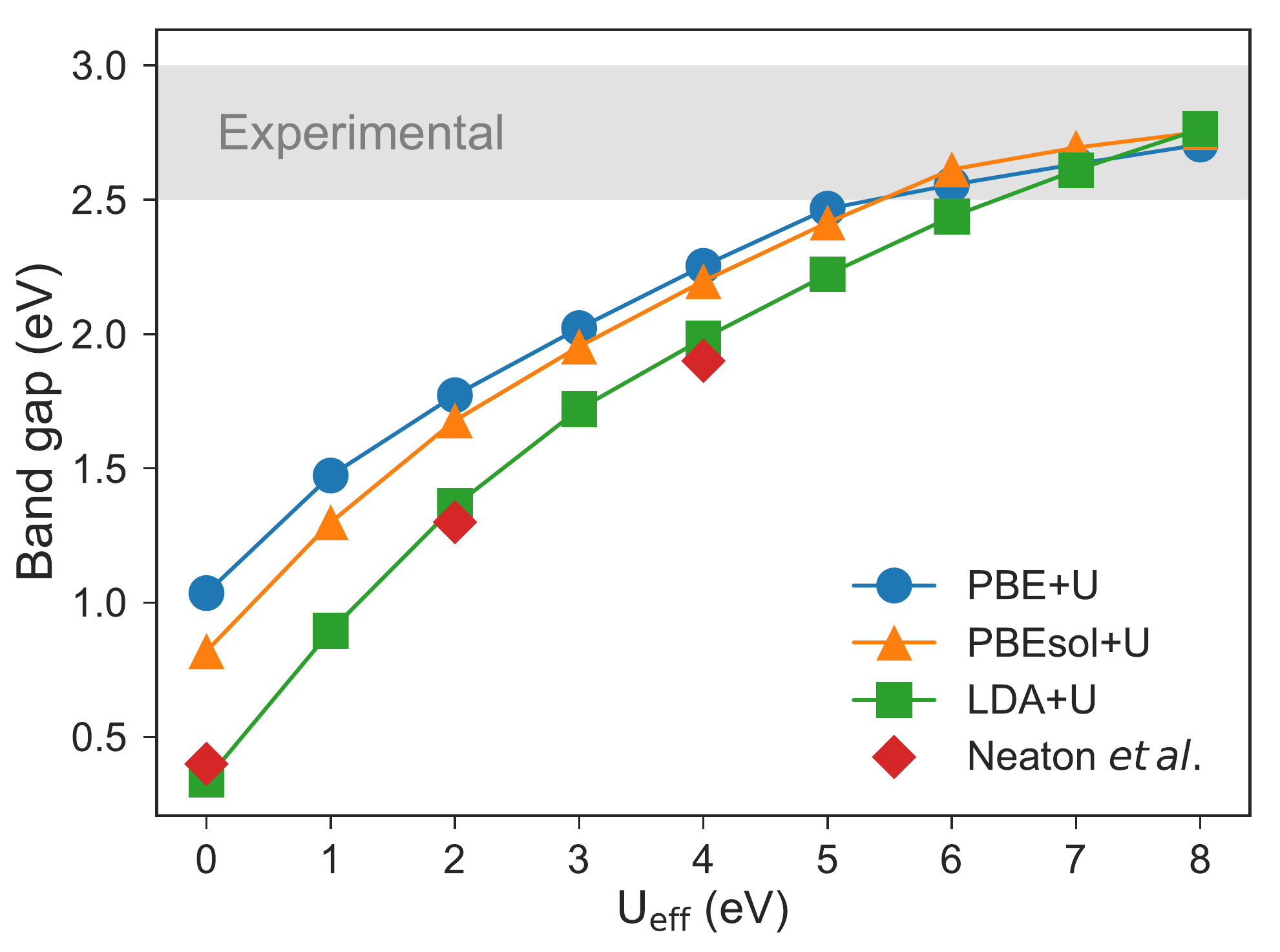}
        \caption{Variation in calculated electronic band gap as a function of $\mathrm{U_{eff}}$ for the PBE+U (blue circles), PBEsol+U (orange triangles) and LDA+U (green squares) xc functionals. The range of band gaps reported in the experimental literature is shown as the shaded region. The LDA+U results reported by Neaton \emph{et al.} (red diamonds) \cite{Neaton2005} are shown for comparison.
        }
        \label{fig:ueff_bandgaps}
    \end{figure}
    % --------------------%

    Given that $\mathrm{U_{eff}}$ is sometimes chosen such that the calculated band gap matches that found in experiment, we now explicitly examine the effect of $\mathrm{U_{eff}}$ on the electronic band gap.
    We expect that, as $\mathrm{U_{eff}}$ increases, the band gap will increase due to the enhanced localisation of the Fe $d$ orbitals.
    Indeed we see such a relationship in Fig. \ref{fig:ueff_bandgaps}, where we plot the electronic band gap against $\mathrm{U_{eff}}$.
    The trends across the PBE+U, PBEsol+U and LDA+U functionals are similar, though the LDA+U gaps are significantly smaller than those of the PBE+U and PBEsol+U functionals for all $\mathrm{U_{eff}}\leq 6$ eV. The LDA+U values are in excellent agreement with those found in Ref.~\cite{Neaton2005}, also plotted in Fig. \ref{fig:ueff_bandgaps}.
    In Ref.~\cite{Young2012a}, the PBE+U band gap is calculated from the theoretical optical absorption spectrum to be 2.58 eV for a $\mathrm{U_{eff}}$ of 5 eV. Their band gap value is 0.11 eV higher than the electronic band gap found in this work, 2.47 eV, for the same $\mathrm{U_{eff}}$ value.

    In order to match the experimental band gap range of 2.5--3.0 eV \cite{Gao2006,Kumar2008,Moubah2012}, a $\mathrm{U_{eff}}$ of 5 eV or larger is clearly required. As we have seen above however, the ordering of the Fe $d$ orbitals at the $\mathrm{CB_{min}}$ inverts for $\mathrm{U_{eff}}>4$ eV, suggesting that fitting $\mathrm{U_{eff}}$ to the electronic band gap alone may introduce some spurious effects.
    While a $\mathrm{U_{eff}}<4$ eV underestimates the electronic band gap, we note that the DFT+U method can, at best, only correct the self-interaction error in the orbitals to which it is applied (in this case the Fe $d$ orbitals). Self-interaction error from the other BFO orbitals, together with other sources of error intrinsic to Kohn-Sham DFT \cite{Perdew1986} are not accounted for by using DFT+U. That is to say, we ought to expect some remaining underestimation of the electronic band gap, even for the value of $\mathrm{U_{eff}}$ that most accurately localises the Fe $d$ orbitals.

%%%%%%%%%%%%%%%%%%%%%%%%%%%%%%%%%%%%%%%%%%%%%%%%%%%%%%%%%%%%%%%%%%%%%%%%%%%%%%

%%%%%%%%%%%%%%%%%%%%%%%%%%%%%%%%%%%%%%%%%%%%%%%%%%%%%%%%%%%%%%%%%%%%%%%%%%%%%%

\section{\label{sec:conclusions}Conclusions\protect}

We have employed the DFT+U method to calculate the optimum crystal geometry and electronic structure of the $R\,3\,c$ phase of BFO for a range of $\mathrm{U_{eff}}$ between 0 and 8 eV, applied to the Fe $d$ orbitals.
We showed that the Bi displacement from its centrosymmetric position, the rotation of the FeO$_6$ octahedra, the distortion of the octahedra, and the spontaneous polarisation change negligibly within the $\mathrm{U_{eff}}$ range typically employed in the context of BFO.

The electronic structure, in contrast, varies significantly with $\mathrm{U_{eff}}$, as designed: the application of $\mathrm{U_{eff}}$ is meant to correct the over-delocalisation of the states to which it is applied. With increasing $\mathrm{U_{eff}}$, we find that the character of the states near the band edges changes, in addition to the band gap, leading to enormous changes in calculated charge carrier effective masses.
In particular, the ordering of the Fe $d$ orbitals at the $\mathrm{CB_{min}}$ inverts for $\mathrm{U_{eff}}$ values larger than 4 eV.

Using a $\mathrm{U_{eff}}$ of 4 eV leads to a 10--25\% underestimation in the calculated band gap with respect to experimental values. To match the experimental band gap, a $\mathrm{U_{eff}}$ value of between 5 and 8 eV would be required. However, in this range of $\mathrm{U_{eff}}$, the $\mathrm{CB_{min}}$ is Fe $e_{g}$ in character rather than the Fe $t_{2g}$ character found for $\mathrm{U_{eff}}$ values less than 5 eV.
The widespread practice of selecting the $\mathrm{U_{eff}}$ parameter to match the experimental band gap therefore clearly needs to be exercised with caution, particularly in cases for which the character of the band edges play a significant role.
We strongly recommend a thorough analysis of the effect of $\mathrm{U_{eff}}$ on the calculated electronic structure before proceeding with calculations that depend on the character of the band edges such as charge carrier effective masses, optical absorption energies and oxidation energies.

%%%%%%%%%%%%%%%%%%%%%%%%%%%%%%%%%%%%%%%%%%%%%%%%%%%%%%%%%%%%%%%%%%%%%%%%%%%%%%

%%%%%%%%%%%%%%%%%%%%%%%%%%%%%%%%%%%%%%%%%%%%%%%%%%%%%%%%%%%%%%%%%%%%%%%%%%%%%%
\section{\label{sec:acknowledgements}Acknowledgements\protect}
The authors acknowledge the use of the UCL Legion (Legion@UCL) and Grace (Grace@UCL) High Performance Computing Facilities, and associated support services, in the completion of this work.

\section{Bibliography}
\label{sec:Bibliography}
\bibliographystyle{iopart-num}
\bibliography{bfo_effecive_mass_paper}

\end{document}